\documentclass[a4paper,12pt]{article}
\usepackage[utf8]{inputenc}
\usepackage{cancel}
\usepackage{ulem}
\usepackage{amsfonts}
\usepackage{amssymb}
\usepackage{graphicx}
\usepackage{amsmath}
\usepackage{enumerate}
\usepackage{subfloat}
\usepackage{subfig}
\usepackage{color}
\usepackage{float}
\usepackage{tikz}
\usepackage{here}
\usepackage{cite}
\usepackage{mathrsfs}
\usepackage{float,epsfig}
\usepackage{dcolumn}
\usepackage{multirow}
\usepackage{placeins}
\usepackage{graphicx}
\usepackage{tikz}
\usepackage{bm}
\usepackage{amsmath,amssymb,amsthm}
\usepackage[colorlinks=true,linkcolor=blue]{hyperref}
\hypersetup{citecolor=magenta}
\title{\bf \Large  
Fingerprinting the fractional order phase transitions  in AdS black holes} 

\author{Mohamed Chabab\footnote{mchabab@uca.ac.ma (Corresponding author)}, and Samir Iraoui\footnote{samir.iraoui@ced.uca.ma}\\
	\\ 
	{\small High Energy and Astrophysics Laboratory, Physics Department, FSSM, Cadi Ayyad University, 
	}\\
	{\small  P.O.B. 2390 Marrakech, Morocco.
	}
}

\date{}

\begin{document} 
\maketitle
	
\begin{abstract}
In this paper, we have extended and deepened the study on fractional order phase transition (FPT) of a charged AdS black hole performed in \cite{Ma:2019hvr}. We have carried out a detailed analysis of FPT for several AdS black hole prototypes: black hole surrounded by quintessence background, $5D$ Gauss-Bonnet, $D$ dimensional  RN-AdS BH, and lastly Kerr black holes. We have
shown that  the $4/3$ order FPT at critical points obtained in \cite{Ma:2019hvr} still holds for the first three black holes systems, while for Kerr black holes, the fractional order is rather $1/3$. These results suggest two remarkable features:  Firstly $4/3$ order phase transition can be assumed for  asymptotically  AdS black holes spherical solutions, secondly the fractional order is not universal and can be affected by the geometric symmetry.  \\

Keywords: AdS black holes, Gauss-Bonnet, Kerr black holes, phase transitions, fractional order.
\end{abstract}

          \tableofcontents
  
\renewcommand{\thefootnote}{\arabic{footnote}}
\setcounter{footnote}{0}
\section{Introduction}
\label{intro}

Thermodynamic properties and particularly phase transitions of black holes in asymptotically Anti de-Sitter spacetime have been investigated extensively. This growing interest in phase transitions originates from  their relation to holographic superconductivity in the context of the AdS/CFT correspondence~\cite{Maldacena}. Recently, many studies of black holes criticality have  identified the cosmological constant with thermodynamic pressure and its conjugate quantity with thermodynamic volume~\cite{Belhaj,Belhaj2,Chabab,Kubiznak}. Furthermore the inclusion of the $P-V$ term in the first law of thermodynamics has led to identification of the  black hole mass with the enthalpy of the event horizon ($H\equiv M$)~\cite{Kastor.2009}.   Then, the analogy with the Van de Waals $P-V$  criticality has been established~\cite{Dolan.2011} and the first and second order phase transitions readily found~\cite{Kubiznak}.  Generalization to thermodynamics structure of Reissner-Nordstr\"om black holes surrounded by quintessence has also been carried out~\cite{Kiselev, Thomas:2012zzc,Li2014, Wu2018meo, Chabab:2020ejk}. \\

Originally outlined by Ehrenfest who used the discontinuities in derivatives of the free energy to probe and classify thermodynamic phase transitions, Ehrenfest's classification scheme is based on the experimental observation of the density contrast between coexisting phases and latent heat during a phase transition. When a system undergoes a phase transition at a critical temperature, Ehrenfest defines its order of phase transition as the smallest integer $n\ge 1$ such that the $n-th$ derivative of $g(T)$ has a discontinuity at the critical point. More formally this criterion generally  reads as:
\begin{equation}
\lim_{ T\rightarrow T_{c}^+}\frac{\mathrm{d} ^n g(T)}{\mathrm{d} T^n}=A^+\ne A^- =  \lim_{ T\rightarrow T_{c}^-}\frac{\mathrm{d} ^n g(T)}{\mathrm{d} T^n},
\end{equation}
 
 \noindent
 where $g(T)$ denotes the free energy as a function of temperature $T$.  \\
 
 Hawking-Page phase transitions is an example of first order transition. The critical point is commonly presented as a witness of second order phase transition in the sense of Ehrenfest. A much more general classification scheme based on fractional derivatives, viz. a derivative with noninteger order, along arbitrary curves in the thermodynamic state space, has been revealed~\cite{Hilfer1991,Hilfer2000,Hilfer1992,Hilfer1992b}.  Fractional phase transitions (FPT) go back to the work of Nagle on dipalmitoyl lecithin (DPL) system~\cite{Nagle},  where it has been observed  that the order$\leftrightarrow$ disorder phase transition is neither of first nor second type,  but is rather of  $3/2$ order (see also~\cite{Agnihotri}). Besides, $4/3$ order FPTs have also been found in Van der Waals fluids~\cite{Hilfer2000}. \\

The generalized Ehrenfest classification holds for all thermodynamic systems having phase structures, including AdS black holes with analogous behaviour to Van der Waals gases. Indeed, a first study  recently performed by Ma~\cite{Ma:2019hvr} showed that RN-AdS black holes can undergo $4/3$ order FPT. In this paper, we aim to explore further this issue and see whether this fractional order holds for any AdS black hole solution, as in VdW systems. To this end, we consider four prototypes: First the charged AdS black hole surrounded by quintessence field to show the effect of   an external source.  Second,  we  use a $D$ dimensional  RN-AdS and $5$ dimensional Gauss-Bonnet-AdS black hole to probe whether higher spacetime dimensions, $d > 4$, and higher derivative corrections can affect and modify the FPT order near the critical point. At last, we also study FPT beyond spherical symmetric static black holes through Kerr black holes, and verify the role played by the geometric symmetry when dealing with FPT. \\

This paper is organized as follows. In the next section, first we present the following AdS black hole solutions: Kiselev, $D$-dimensional RN-AdS, and $5D$ Gauss-Bonnet solution, and their corresponding thermodynamic quantities. Then,  we study their fractional phase transitions at critical points via the Caputo derivative of Gibbs free energy. Section~\ref{Kerr} is devoted to the analysis of Kerr-AdS black hole and derivation of its FPT order. Our conclusion is drawn in the last section.

 \section{Thermodynamic criticality and fractional order phase transitions   of  AdS black holes with spherical symmetry}
\label{Thermo}
In this section, we consider  three spherical symmetric solutions of static asymptotically  AdS black holes, namely: a charged-AdS BH surrounded by quintessence, $D$ dimensional RN-AdS BH and a $5$D Gauss-Bonnet AdS black holes. We first calculate their thermodynamic quantities,  then describe the behaviors of corresponding Gibbs free energy and their fractional derivatives nearby the critical points. Our objective is to probe whether the extra factors, as quintessence parameter, higher dimensions and/or higher derivatives in Gauss-Bonnet term parameters, could affect the FPT order.

\subsection{Charged AdS black hole surrounded by quintessence}
 From high precision astronomical observations, it  has been shown that the universe is currently undergoing a phase of accelerated expansion~\cite{Riess,Perlmutter}, which might be due to dark energy acting as a repulsive gravity. A possible origin of this phenomena could come from the so-called quintessence field,  which obeys an equation of state formulated through the relation between negative pressure and  energy density as, $p=\omega_q \rho_{q}$, where the quintessence parameter is constrained as: $-1<\omega<-1/3$~\cite{Caldwell}.
 
 We consider the Kiselev solution of four dimensional charged AdS black holes surrounded by quintessence,   
\begin{equation}\label{Kilesev}
f(r)=1-\frac{2 M}{r}+\frac{Q^{2}}{r^{2}}-\frac{ \Lambda}{3}r^{2}-\frac{\alpha}{r^{3\omega_{q}+1}},
\end{equation} 
where $\alpha$ represents a positive normalisation parameter, while $M$ and $Q$ are the mass  and  electric charge of the black hole respectively. As usual, we treat the cosmological constant as a dynamical pressure of the black hole~\cite{Kastor.2009},
\begin{equation}\label{pressurecosmolog}
P=-\frac{\Lambda}{8\pi}.
\end{equation}
 The Hawking temperature  related to the surface gravity via the formula $2\pi T=\kappa$,  is then expressed as~\cite{Chabab:2020ejk},
\begin{equation}\label{temperature}
T=\frac{1}{4\pi }\left[\frac{1}{   r_h} - \frac{Q^2}{r_h^{3}}+\frac{  3
	\omega_q \alpha}{ r_h^{3 \omega_q  +2}}+8 \pi   r_h P\right],
\end{equation}
where the horizon $r_{h}$ is determined from the condition $f\left(r_{h}\right)=0$. 
The equation of state for the charged AdS black hole surrounded by quintessence is given by:
\begin{equation}
P=\frac{2 \pi T}{2 \pi v}-\frac{1}{2\pi v^2}+\frac{2Q^2}{\pi v^4}-\frac{3\times 2^{1+3\omega_q} \alpha \omega_q}{2 \pi v^{3 \omega_q +3}},
\end{equation}
where the specific volume $v$ associated  with the fluid volume is related to the horizon radius since $v=2 r_{h}$.  The first law of black hole
thermodynamics in the extended phase space can be written as~\cite{Li2014}, \begin{equation}\label{firstlaw}
dM=T\mathrm{d}S+V\mathrm{d}P+\Phi \mathrm{d}Q+\mathcal{Q} \mathrm{d}\alpha,
\end{equation}
where the conjugate quantities of the  parameters $P$, $Q$ and $\alpha$ read respectively as
\begin{align}\label{conjugatequantities}
V =  \frac{4\pi r_h^{3}}{3},\quad \Phi  = \frac{Q}{r_h},\quad \mathcal{Q} = -\frac{1}{2 r_{h}^{3 \omega_q }},
\end{align}
while the Smarr relation is formulated by,
\begin{equation}\label{Smarrformula}
M=2 T S-2  P V+\Phi Q+\left(3 \omega_q+1\right) \mathcal{Q} \alpha.
\end{equation}

Thereafter, without loss of generality, we will  set $\omega_{q}$ to the value $\omega_{q}=-2/3$. In this case,  we can determine  the critical point analytically as,
\begin{equation}
T_c=\frac{\sqrt{6}-9 Q \alpha}{18\pi Q}, \quad v_c=2\sqrt{6}Q, \quad P_c=\frac{1}{96\pi Q^2}.
\end{equation}
For subsequent analysis, it is more appropriate to introduce  the following variables,
\begin{equation}\label{dimpvt}
p=\frac{P-P_c}{P_c}, \qquad t=\frac{T-T_c}{T_c},  \qquad \nu=\frac{v-v_c}{v_c}.
\end{equation}
Thus with the new set of variables $({t}, {p}, {\nu})$, the critical point lies at $({t}={p}={\nu}=0)$, while  the equation of state reduces to a quartic equation in $\nu$, 
\begin{equation}\label{eosbh}
3 p (\nu+1)^4+4 t (\nu+1)^3 \left(3 \sqrt{6}   Q \alpha -2\right)+\nu^3 (3 \nu+4)=0,
\end{equation}
Besides, the rescaled Gibbs free energy  now takes  the form,
\begin{equation}g(t,p)=\frac{8-\nu ^4-4 \nu ^3+8 \nu -(\nu +1)^4 p}{4 \sqrt{6} (\nu +1)}.
\end{equation}

Next we probe  the behavior of Gibbs free energy $g(t,p)$ and its fractional derivatives near critical point. First we use the expansion series of $ \nu(t, p)$ derived from Eq.~\eqref{eosbh} and substitute it into the  $g(t,p)$. The resulting expression reads as,
\begin{multline}\label{gRN}
g(t,p)\approx\left[\sqrt{\frac{2}{3}}+\frac{p}{2 \sqrt{6}}+\dots\right]
 -\left[\sqrt{\frac{2}{3}}-3 Q \alpha
-\frac{\left(2-3 \sqrt{6}   Q \alpha\right) }{6^{1/6}}\left(p^{1/3}-\frac{19 }{4\times 6^{2/3}}p^{2/3}\right)+\dots\right]t \nonumber \\
 -\left[ \frac{2^{1/6}\left(2-6 \sqrt{6} Q\alpha+27 Q^{2}\alpha^{2} \right)}{9 \times 3^{1/6}} \left(\frac{ 2^{5/3}   }{p^{2/3}}-\frac{13 }{ 9^{1/3} p^{1/3}}\right)+\dots\right]t^2 + \mathcal{O}[ t^3].
\end{multline}
 The above equation can well  be expressed in the Taylor series of $t$, and written in a simple form as,
\begin{equation}
g(t, p)=A( p)+B( p) t+D( p) t^2+\mathcal{O}[ t^3].
\end{equation}

Several definitions of  fractional derivatives exists in literature~\cite{Pod1999}. Here we rely on Caputo definition which enables easy use of conventional boundaries and initial conditions~\cite{1967GeoJ}:
\begin{equation}
D_{ t}^{\beta}g( t)=\frac{1}{\Gamma\left(n-\beta\right)}\int_{0}^{t}\left(t-\tau\right)^{n-\beta-1}\, \frac{\partial^{n}g(\tau)}{\partial \tau^{n}}\mathrm{d}\tau,\qquad  n-1<\beta<n,
\end{equation}
where $\beta$ is the order of derivative and  $n$  an integer. As a result of calculation we get:
\begin{align}\label{fractionalgibbs}
\left.D_{ t}^{\beta}g( t, p)\right|_{ p}&=&
\begin{cases}
\dfrac{2 D(p) }{\Gamma\left(3-\beta\right)}t^{2-\beta}+\dfrac{\left(\beta-2\right) B(p)  }{\Gamma\left(3-\beta\right)}t^{1-\beta}, & t>0;\ 0<\beta<1,\\ 
-\dfrac{2 D(p) }{\Gamma\left(3-\beta\right)}\left|t\right|^{2-\beta}-\dfrac{\left(\beta-2\right) B(p)   }{\Gamma\left(3-\beta\right)}\left|t\right|^{1-\beta},& t<0;\ 0<\beta<1,\\ 
\dfrac{2 D(p)}{\Gamma\left(3-\beta\right)}t^{2-\beta},& t>0;\ 1<\beta<2,\\ 
-\dfrac{2 D(p)}{\Gamma\left(3-\beta\right)}\left|t\right|^{2-\beta},& t<0;\ 1<\beta<2.
\end{cases}
\end{align}

 According to~\cite{Li2014,Chabab:2020ejk}, the equation of state near  the critical point ($t\ll 1,\ \nu\ll 1$) simplifies to:
\begin{equation}\label{vdwpt}
p \approx k t +O[t^{2}, t \nu],
\end{equation}
where the slope $k$  is given by,
\begin{equation}\label{knearcrit}
k=\frac{T_c}{P_c v_c}=\frac{8}{3}-4\sqrt{6} Q\alpha.
\end{equation}
Therefore, we can calculate the values of $D_{ t}^{\beta}g( t, p)$ for $1<\beta<2$ in the limit $( t \rightarrow 0,~ p\rightarrow 0)$ by substituting Eq.~\eqref{vdwpt} into Eq.~\eqref{fractionalgibbs}. Thus, we  obtain
\begin{equation}
\lim_{ t\rightarrow 0^{\pm}}D_{ t}^{\beta}g( t, p)=\left\{
\begin{array}{lr}
	0 ~~&\text{for}~~\beta<4/3,\\
	\mp\frac{\left(2-3 \sqrt{6} \alpha  Q\right)^{4/3}}{\sqrt{6} \Gamma \left(\frac{2}{3}\right)} &\text{for}~~\beta=4/3,\\
	\mp\infty ~~ &\text{for}~~\beta>4/3.
\end{array}
\right.\end{equation}

\begin{figure}[!h]
	\begin{center}
				\hspace*{-2em}
		\begin{tikzpicture}[remember picture]
		\node[anchor=south west,inner sep=0] (image) at (0,0) {\includegraphics[height=5.4cm]{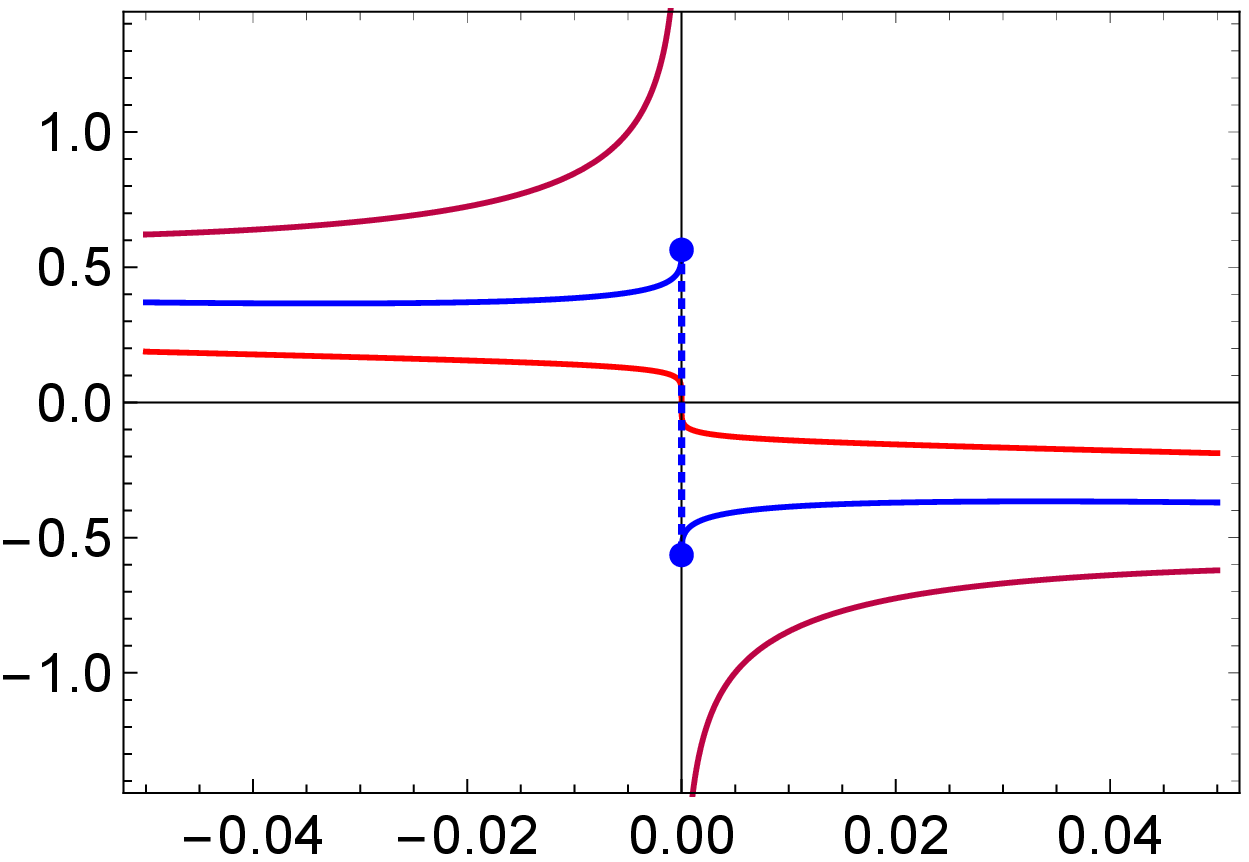}};
		\begin{scope}[x={(image.south east)},y={(image.north west)}]
		\node[font=\boldmath] at (.34, .81) {\footnotesize \bfseries \color{purple} $\beta=9/6$ };
		\node[font=\boldmath] at (.34,.68) { \footnotesize \bfseries \color{blue} $\beta=4/3$ };
		\node[font=\boldmath] at (.34,.54) { \footnotesize \bfseries \color{red} $\beta=9/8$ };
		\node[font=\boldmath] at (-.03,.51) { \footnotesize \bfseries $D_{ t}^{\beta}g$ };
		\node[font=\boldmath] at (.55,-.04) { \footnotesize \bfseries  $t$ };
		\end{scope}
		\end{tikzpicture}
		\hspace*{.5em}
		\begin{tikzpicture}[remember picture]
		\node[anchor=south west,inner sep=0] (image) at (0,0) {\includegraphics[height=5.32cm]{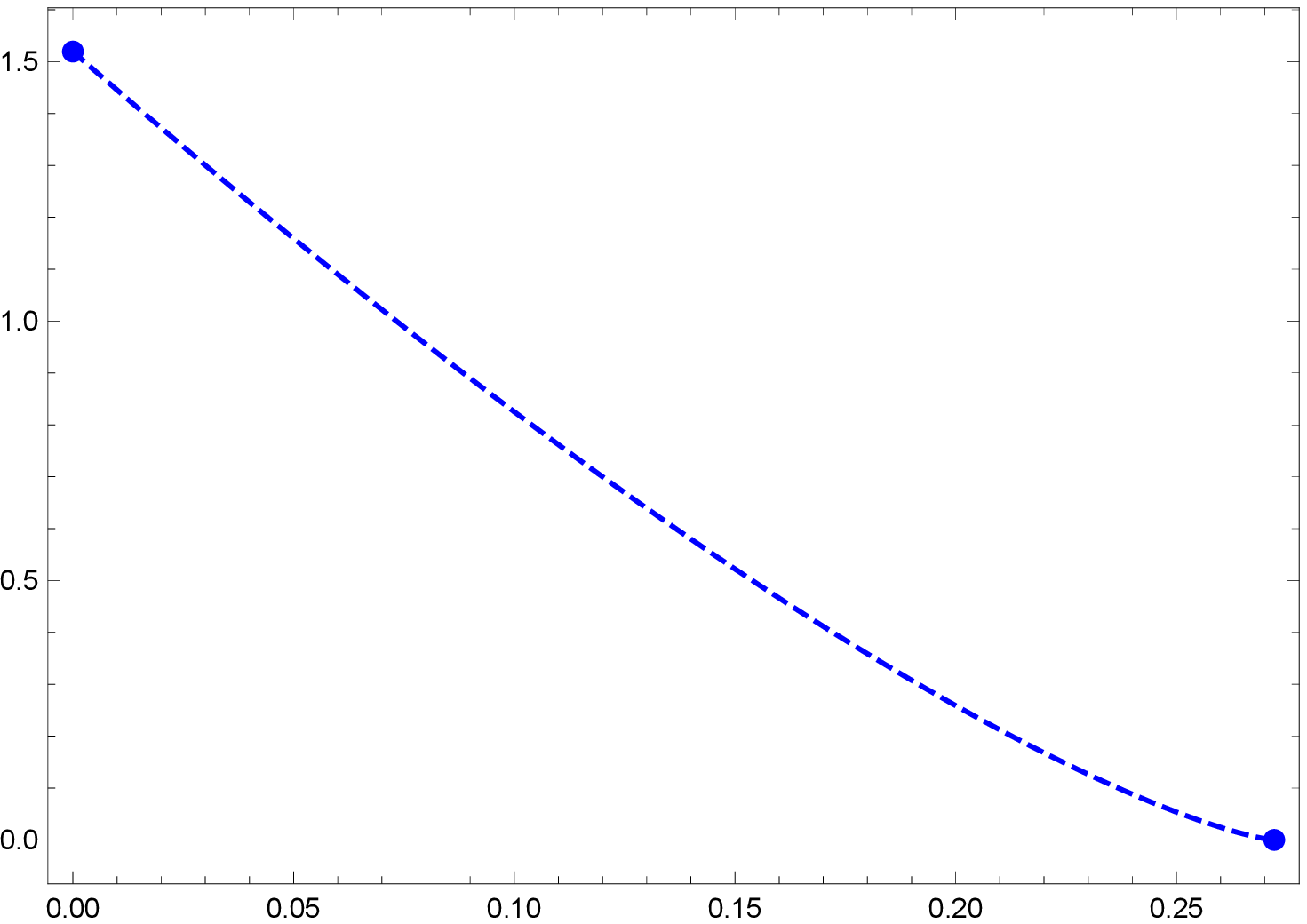}};
		\begin{scope}[x={(image.south east)},y={(image.north west)}]
		\node at (.105, .5) {\footnotesize \bfseries  \rotatebox{90}{Jump magnitude} };
		\node[font=\boldmath] at (.55,-.04) { \footnotesize \bfseries  $\alpha \times Q$ };
		\end{scope}
		\end{tikzpicture}
	\end{center}
	\caption{Left panel: The behavior of  fractional derivatives  $D_{ t}^{\beta}g$ for $\beta$ equals to: $9/6$ (purple line), $9/8$ (red line) and $4/3$ (blue line) near the critical point of charged AdS black hole surrounded by quintessence. Input: $Q \alpha=0.55$. Right panel: Jump magnitude near the critical point as a function of $\alpha$ for  $\beta=4/3$. }
	\label{figrndg}
\end{figure}

Obviously, for  $\beta=4/3$ case, we see a jump discontinuity,
\begin{equation}
\lim_{ t\rightarrow 0^{-}}D_{ t}^{\beta}g \neq \lim_{ t\rightarrow 0^{+}}D_{ t}^{\beta}g.
\end{equation}
When $\beta > 4/3$, the $\beta$-order fractional derivatives of the Gibbs free energy diverge. Hence, as for $4D$ RN-AdS black holes~\cite{Ma:2019hvr}, the FPT order of the RN-AdS Black hole surrounded by quintessence is $\beta=4/3$.  This is clearly illustrated by the left panel of  Fig.~\ref{figrndg} where the fractional derivative $D_{ t}^{\beta}g$  is plotted as a function of the reduced temperature $t$ for several values of $\beta$: $\beta=9/6$, $9/8$  and $4/3$  near the critical point. One can see that this plot is continuous at $T_c$ for $\beta < 4/3$, becomes progressively steeper as $\beta$ approaches $4/3$, but behaves as discontinuous at $T_c $ for $\beta = 4/3$ and fully divergent once $\beta > 4/3$. It is also worth to notice the behaviour of the magnitude $\frac{\left(4-6 \sqrt{6} \alpha  Q\right)^{4/3}}{\sqrt{6} \Gamma \left(\frac{2}{3}\right)}$ for $\beta = 4/3$ shown in Eq.~$18$. This magnitude monotonically decreases as far as $\alpha$ increases and becomes $\frac{4^{4/3}}{\sqrt{6} \Gamma \left(\frac{2}{3}\right)}$ when $\alpha\rightarrow 0,$ whereas it fades away in the limit $\alpha\rightarrow \sqrt{6}/(9 Q)$, i.e, $T_c\rightarrow 0$, as shown by the right panel of Fig.~\ref{figrndg}. \\

This result clearly shows that, for spherical symmetric AdS black hole in quintessence background,  the FPT still stand at $4/3$ order. Therefore, the fractional phase transition is not affected by the external quintessence field surrounding the charged AdS black hole. 

	
	\subsection{Charged AdS black holes in higher dimensions }
	 Now we consider a $D$ dimensional  charged AdS  black holes to check whether  higher dimension $D$ deflects the fractional order of the phase transition from $4/3$. For this case, the metric is given by, 
	\begin{align}
		ds^2 &= -f \mathrm{d}t^2 + \frac{\mathrm{d}r^2}{f} + r^{2} \mathrm{d}\Omega_{D-2}^2,\\
		f(r) &=1-\frac{m}{r^{D-3}} +\frac{q^2}{r^{2(D-3)}}    -\frac{2 \Lambda  r^2}{(D-2) (D-1)}.
	\end{align}
	$m$ and $q$ are related to the ADM mass $M$ while $Q$ stands for the black hole charge~\cite{Chamblin1999tk}:
	\begin{align}
		M=\frac{D-2}{16 \pi} \omega_{D-2} m,\qquad Q=\frac{\sqrt{2(D-2)(D-3)}}{8\pi}\,\omega_{D-2}\,q,
	\end{align}
	with $\omega_D$ is the volume of the unit $D$-sphere.
	
	Thermodynamics of higher dimension black holes has been investigated in ~\cite{Gunasekaran2012dq} as well as their critical behaviors revealed as phase transition between small and large black holes. The critical points were derived at, 
	\begin{align}
		v_c & = \frac{4}{D-2}\left[q^2 (D-2)(2D-5) \right]^{\frac{1}{2(D-3)}}, \nonumber\\
		T_c &= \frac{(D-2)(d-3)^2}{  (2D-5)}\frac{1}{4\pi  v_c},\nonumber\\
		P_c &= \left(\frac{D-3}{D-2 }\right)^2\frac{1}{\pi  {v_c}^2},
	\end{align}
	where the  specific volume is given, in the geometric units, by $v=\frac{4}{D-2} r_h$. As previously once we use the reduced variable, the equation of state transforms to the following general form, 
	\begin{align}\label{treduced}
		4 t=\frac{2 D-5}{(D-3) (\nu +1)}-\frac{(\nu +1)^{5-2 D}}{6+D(D-5)}+\frac{D (2 D-5) (\nu +1)
			(p+1)}{(D-2)^2}-\frac{2 (2 D-5) (\nu +1) (p+1)}{(D-2)^2}-4.
	\end{align}
	The rescaled Gibbs free energy is
	\begin{align}
		g(t,p)=\frac{(5-2 D)^2 (\nu +1)^{3-D}}{2 [(D-2) (2 D-5)]^{3/2}}+\frac{1}{2} \sqrt{\frac{2 D-5}{D-2}} (\nu +1)^{D-3}-\sqrt{\frac{2 D-5}{(D-2)^3}}\, \frac{(D-3)^2  }{2
			(D-1)}(p+1) (\nu +1)^{D-1},
	\end{align}
	Note that for $D=4$ the results obtained in~\cite{Ma:2019hvr} for RN-AdS$_4$ BH is recovered.
	
	Next step, we solve equation~\eqref{treduced} and expand the Gibbs free energy  for each dimension spacetime $D>5$. As illustration, consider $D=6$, then the    equation for $g(t,p)$ near the critical point reduced to, 
	\begin{multline}
		g(t,p)
		\approx
		\left(\frac{\sqrt{7}}{5}+\frac{9 \sqrt{7} p}{20}+\dots\right)+
		\left(-\frac{9}{\sqrt{7}}+\frac{18 \sqrt[3]{3} p^{1/3}}{\sqrt{7}}-\frac{15}{4}
		\left(3^{2/3} \sqrt{7}\right) p^{2/3}+\frac{60677 p}{768
			\sqrt{7}}+\dots\right)\, t\\
		+\left(-\frac{48 \sqrt[3]{3}}{7 \sqrt{7}
			p^{2/3}}+\frac{116\ 3^{2/3}}{7 \sqrt{7} p^{1/3}}+\dots\right)\, t^2 
		+\mathcal{O}[ t^3].
	\end{multline}
	Since the equation of state around  near the critical point  behaves as  $p\approx \frac{4D-8}{2D-5} t$, and for $t$ approaching $0$, we find  that the fractional derivative of $g(t,p)$ is discontinuous function, 
	\begin{equation}
		\lim_{ t\rightarrow 0^{\pm}}D_{ t}^{\beta}g( t, p)=\left\{
		\begin{array}{lr}
			0 ~~&\text{for}~~\beta<4/3,\\
			\mp\dfrac{12 \sqrt[3]{6}}{7^{5/6} \, \Gamma \left(\frac{5}{3}\right)}  &\text{for}~~\beta=4/3,\\
			\mp\infty ~~ &\text{for}~~\beta>4/3,
		\end{array}
		\right.\end{equation} 
	with a discontinuity taking place at that $\beta=\frac{4}{3}$.  We summarized the  calculation results for all dimensions $D=4 -10$ in Table~\ref{tableRNAdS}.
	\begin{table}[!h]
		\begin{center}
			\begin{tabular}{ccc}
				\hline 
				Dimension $d$& order FPT  &  $\lim_{ t\rightarrow 0^{\pm}}D_{ t}^{\beta}g( t, p)$ \\  
				\hline
				$D=4$	& $\beta=4/3$  & $\mp \dfrac{2\ 2^{5/6}}{3 \sqrt{3}\, \Gamma \left(\frac{5}{3}\right)}$ \\ 
				$D=5$	& $\beta=4/3$  & $\mp \dfrac{8 \sqrt[3]{2}}{\sqrt[6]{3}\, 5^{5/6}\, \Gamma \left(\frac{5}{3}\right)}$ \\  
				$D=6$	& $\beta=4/3$  &  $\mp  \dfrac{12 \sqrt[3]{6}}{7^{5/6} \, \Gamma \left(\frac{5}{3}\right)}  $ \\  
				$D=7$	& $\beta=4/3$  & $\mp  \dfrac{32 \sqrt[3]{\frac{2}{3}} \sqrt{5}}{9\, \Gamma \left(\frac{5}{3}\right)}  $ \\ 
				$D=8$	& $\beta=4/3$  &  $\mp  \dfrac{50 \left(\frac{2}{11}\right)^{5/6}}{\sqrt[6]{3}\, \Gamma \left(\frac{5}{3}\right)}  $\\ 
				$D=9$	& $\beta=4/3$  &  $\mp  \dfrac{24 \sqrt[3]{6} \sqrt{7}}{13^{5/6} \, \Gamma \left(\frac{5}{3}\right)}  $\\ 
				$D=10$	& $\beta=4/3$  & $\mp  \frac{196 \left(\frac{2}{5}\right)^{5/6}}{3 \sqrt{3} \, \Gamma \left(\frac{5}{3}\right)}  $ \\ 
				\hline 
			\end{tabular} 
			\caption{Limits and fractional order phase transition at the critical point for $D$ dimensional RN-AdS BH.}
			\label{tableRNAdS}
		\end{center}
	\end{table}

	Table $1$ decidedly shows  that, whatever the dimension $D$ of  the RN-AdS black hole, the fractional  order of phase transition near the critical point arises decidedly at $\beta=\frac{5}{3}$,   with the jump magnitude increases as $D$ dimension gets larger.

\subsection{5D Gauss-Bonnet-AdS black hole}
 First we briefly introduce the action of Gauss-Bonnet Black hole and its main thermodynamic features.  \\
 
 The action of this theory reads as:
\begin{equation}\label{actionGaussBonnet}
\mathcal{S}_{L}=\int \mathrm{d}^{D}x \sqrt{-g} \left(R-2\Lambda+\alpha'\mathcal{L}_{2}\right),
\end{equation} 
  $\mathcal{L}_{2}$ represents the Gauss-Bonnet term $\mathcal{L}_{2}= R_{\mu\nu\lambda\sigma} R^{\mu\nu\lambda\sigma}-4R_{\mu\nu} R^{\mu\nu}+R^{2}$,   with the Riemann curvature tensor $R_{\mu\nu\lambda\sigma}$, Ricci tensor $R_{\mu\nu}$, while  $R$ is the Ricci scalar. $\alpha'$ is a dimensionless  coupling constant.
 
A spherically symmetric static solution of this theory has been derived, and  the metric function determined through solving  the real roots of a polynomial equation~\cite{Wheeler1985nh, Wiltshire}.

It is founded to seek a connection between the higher derivatives  in Gauss-Bonnet term  and the phase transition, checking whether it affects the FPT's order. Here, we use as prototype a $D=5$ Gauss-Bonnet-AdS black holes, which exhibit a critical behaviour featured via small/large  phase transition, while for $D \ge 6 $ no phase structure has been unveiled~\cite{Cai2013qga, Belhaj:2014eha}.  The exact solution describing this black hole is given by~\cite{Wiltshire}

\begin{equation}\label{fGB}
f_{GB}(r)=1+\frac{r^{2}}{2 \tilde{\alpha}}\left(1-\sqrt{1-\frac{16 \pi P\tilde{\alpha}}{3}+\frac{32 M \tilde{\alpha}}{3 \pi r^{4}}}\right),
\end{equation} 
where $\tilde{\alpha}$ is a  Gauss-Bonnet coupling constant. In the extended phase space, the equation of states reads as:
\begin{equation}
P=\frac{32 \tilde{\alpha}  T}{9 v^3}+\frac{T}{v}-\frac{2}{3 \pi  v^2},
\end{equation}
where the specific volume is $v=\frac{4}{3} r_h$. The thermodynamic critical point coordinate are
\begin{equation}
P_c=\frac{1}{48 \pi \tilde{\alpha} },\quad T_c=\frac{1}{2 \sqrt{6} \pi \sqrt{\tilde{\alpha}} },\quad v_c=4\sqrt{\frac{2 \tilde{\alpha}}{3  }},
\end{equation}
while the rescaled Gibbs free energy ($G/\tilde{\alpha}$) reads as,
\begin{equation}
g(t,p)=\frac{\nu ^3 (\nu +2)^3+(\nu  (\nu +2)+4) (\nu +1)^4 p}{8 (3 \nu  (\nu +2)+4)}.
\end{equation}
As previous analysis, since the equation of state near  the critical point~\cite{Cai2013qga} is reduced to, $p \approx 4 t +O[t^{2}, t \nu]$, thus the fractional derivative of $g$ can be written as the following form:
\begin{equation}
\lim_{ t\rightarrow 0^{\pm}}D_{ t}^{\beta}g( t, p)=\left\{
\begin{array}{lr}
0 ~~&\text{for}~~\beta<4/3,\\
\mp\dfrac{2^{5/3} }{ \Gamma\left(\frac{5}{3}\right)}\pi &\text{for}~~\beta=4/3,\\
\mp\infty ~~ &\text{for}~~\beta>4/3.
\end{array}
\right.\end{equation}

Again, we find that the phase transition near the critical point is fractional  and happens at order 4/3,  which suggest that neither the spacetime dimension nor the higher derivative corrections introduced by means of the Gauss-Bonnet term can alter the fractional order  phase transition at the critical point. Hence we assume that FPT is located at $4/3$ order , whatever the external factors ( space dimension, surrounding background) as far as the AdS black hole solution possesses a spherical symmetry. \\

The next legitimate question is what about axisymmetric black hole solutions? Is the $4/3$ order of FPT universal? we  attempt to respond to this issue in the subsequent section using Kerr-AdS black hole as prototype.

 \section{Kerr-AdS black holes}
 \label{Kerr}
The Kerr solution is the only known family of exact solutions which could present the stationary axisymmetric field outside a rotating massive object. 
	
By using Boyer-Lindquist coordinates, the Kerr asymptotically AdS black hole solutions can read as~\cite{Kerr1963ud}:
 \begin{align}
 	ds^2=-\frac{\Delta}{\rho^2}\left[dt-\frac{a\sin^2\!\theta}{\Xi}d\varphi\right]^2 +\frac{\rho^2}{\Delta} dr^2+\frac{\rho^2}{S}d\theta^2
 	+\frac{S\sin^2\!\theta}{\rho^2}\left[a dt-\frac{r^2+a^2}{\Xi}d\varphi\right]^2\,,
 \end{align}
where
\begin{align}
	\rho^2=r^2+a^2\cos^2\!\theta\,, \Xi=1-\frac{a^2}{l^2}\,,S=1-\frac{a^2}{l^2}\cos^2\!\theta\,,\Delta=(r^2+a^2)\left(1+\frac{r^2}{l^2}\right)-2mr\,.
\end{align}
 The mass $M$ and the angular momentum $J$ are related to the parameters $m$ and $a$ as follows,
\begin{align}\label{physical}
M=\frac{m}{\Xi^2}\,,\quad J=\frac{am}{\Xi^2}\,.
\end{align}
Knowing that the pressure is still given by Eq.~\eqref{pressurecosmolog} and neglecting all terms higher order in $J$, the equation of state in terms of temperature, specific volume and the angular momentum reds as~\cite{Gunasekaran2012dq}
\begin{align}
P=\frac{T}{v}-\frac{1}{2\pi v^2}+\frac{48 }{\pi v^6}J^2+\mathcal{O}[J^4],
\end{align}
where the specific volume associated to the Kerr black hole is given by:
\begin{align}
v=2\left(\frac{3V}{4\pi}\right)^{1/3}\!\!\!\!
=2r_++\frac{12}{r_+^3(3+8\pi r_+^2 P)} J^2\,.
\end{align}
 The critical point occurs at
 \begin{align}
v_c=2\times 90^{1/4}\sqrt{J}\,,\quad
T_c=\frac{90^{3/4}}{225\pi} \frac{1}{\sqrt{J}}\,,\quad
P_c=\frac{1}{12\sqrt{90} \pi}\frac{1}{J} .
 \end{align}

From the  approximation used previously, we derive the expression of rescaled Gibbs free energy  in terms of the dimensionless parameters given in Eq.~\eqref{dimpvt},
\begin{align}\label{Gibbsrescaled}
g(t,p)=G/\sqrt{J}=\frac{\sqrt[4]{5} \left(-12 \sqrt[4]{10} \sqrt{J} (\nu +1)^7 (p+1)+9 \sqrt{3} (\nu
	+1)^4+\sqrt{3}\right)}{18\ 2^{3/4} (\nu +1)^3}.
\end{align}
and the reduced equation of state:
\begin{align}\label{eqsattreduced}
\nu ^3 \left(10 \nu ^3+36 \nu ^2+45 \nu +20\right)+10 (\nu +1)^6 p-24 (\nu +1)^5 t=0.
\end{align}

Then, using Eqs.~\eqref{Gibbsrescaled} and \eqref{eqsattreduced}, one can readily  derive the phase transition order near the critical point. Indeed,  t-expansion of the rescaled Gibbs free energy becomes, 
\begin{multline}\label{gKerr}
g(t,p)\approx C_1
+\left[C_2+\frac{C_3}{p^{2/3}}+\frac{C_4}{p^{1/3}}+\dots\right]t  
+\left[C_5+\frac{C_6}{p^{5/3}}+\frac{C_7}{p^{4/3}}+\frac{C_8}{p^{2/3}}++\frac{C_9}{p^{1/3}}+\dots\right]t^2 + \mathcal{O}[ t^3],
\end{multline}
where the coefficients $ C_i $ are  functions of $ J $. Moreover, the pressure  behaves as $p\approx \frac{12}{5} t$ near the critical point. Using Eq.~\eqref{fractionalgibbs}, we obtain the fractional derivative of $g(t,p)$ with $0<\beta<2$  in the limits $t \rightarrow 0$ and $p\rightarrow 0$,

\begin{equation}
\lim_{ t\rightarrow 0^{\pm}}D_{ t}^{\beta}g( t, p)=\left\{
\begin{array}{lr}
0 ~~&\text{for}~~\beta<1/3,\\
\pm\dfrac{5 \left(\frac{10}{3}\right)^{2/3} \left(2C_3-C_6\right)}{9 \Gamma \left(\frac{11}{3}\right)} &\text{for}~~\beta=1/3,\\
\mp\infty ~~ &\text{for}~~\beta>1/3,
\end{array}
\right.\end{equation}
with, 
\begin{equation}
C_3=\frac{\sqrt[6]{2} \left(2^{3/4} \sqrt{3}-16 \sqrt[4]{5} \sqrt{J}\right)}{3\times 5^{3/4}};\qquad C_6=\frac{4 \sqrt[6]{2} \left(2^{3/4} \sqrt{3}-16 \sqrt[4]{5} \sqrt{J}\right)}{15\times
	5^{3/4}}.
\end{equation}

\begin{figure}[!h]
 	\begin{center}
 		\begin{tikzpicture}[remember picture]
 		\node[anchor=south west,inner sep=0] (image) at (0,0) {\includegraphics[width=.6\linewidth]{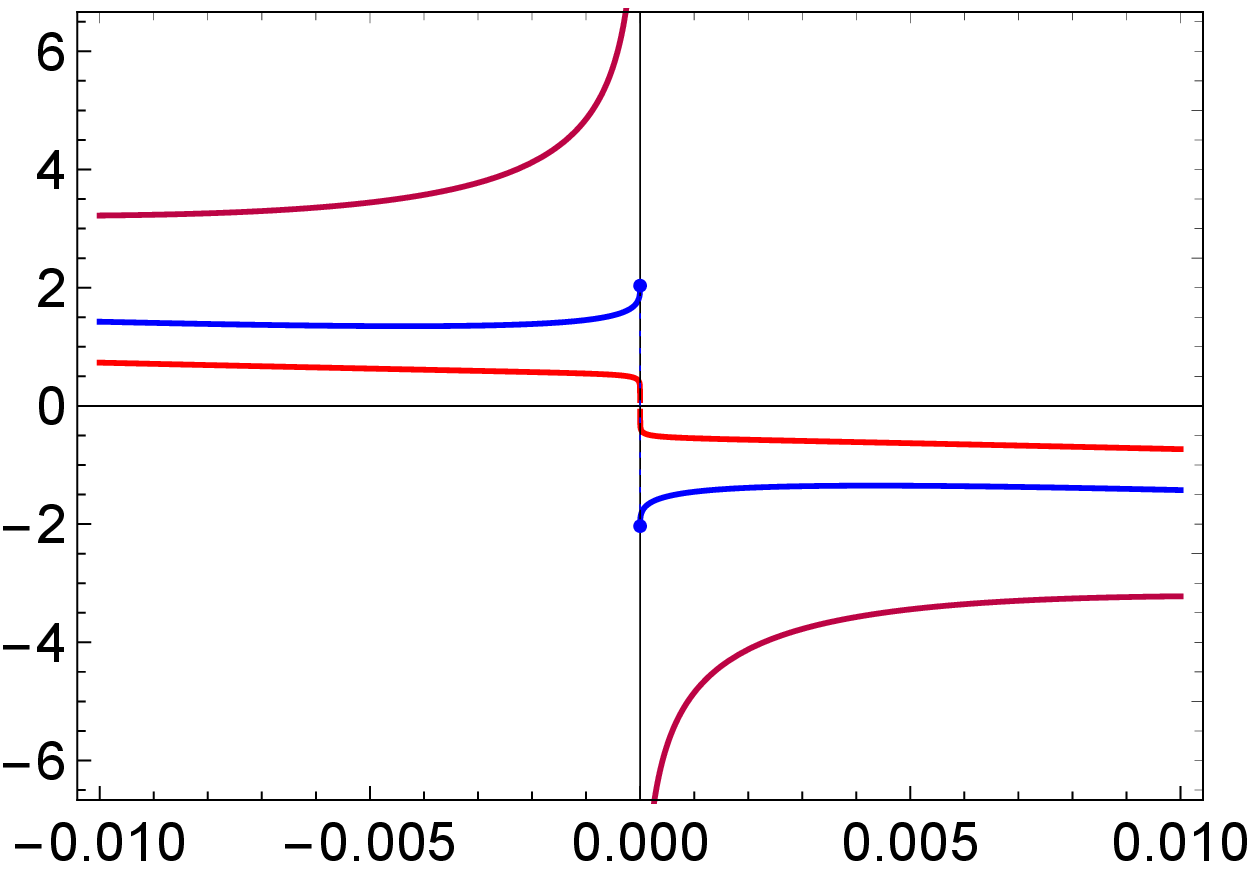}};
 		\begin{scope}[x={(image.south east)},y={(image.north west)}]
 		\node[font=\boldmath] at (.3, .81) {\footnotesize \bfseries \color{purple} $\beta=1/2$ };
 		\node[font=\boldmath] at (.3,.65) { \footnotesize \bfseries \color{blue} $\beta=1/3$ };
 		\node[font=\boldmath] at (.3,.54) { \footnotesize \bfseries \color{red} $\beta=1/6$ };
 		\node[font=\boldmath] at (-.03,.51) { \footnotesize \bfseries $D_{ t}^{\beta}g$ };
 		\node[font=\boldmath] at (.55,-.03) { \footnotesize \bfseries  $t$ };
 		\end{scope}
 		\end{tikzpicture}
 	\end{center}
 	\caption{The behaviors of fractional derivatives  $D_{ t}^{\beta}g$ for $\beta$ equal to: $1/2$ (purple line), $1/6$ (red line) and $1/3$ (blue line) near the critical point of Kerr-AdS black hole, with $J=1$.}\label{3}
 \end{figure}

 The fractional order phase transition of Kerr AdS black hole  clearly arises at $\beta=1/3$.   Figure~\ref{3} illustrates the fractional derivative $D_{ t}^{\beta}g$  for $\beta$ equals to: $1/2, 1/6$  and $1/3$  near the critical point. 

Therefore, near the critical point, we see that fractional order of the phase transition is no longer $\beta=4/3$. This may suggest $4/3$ order FPT is not universal and only holds for static black holes with spherical symmetry.

\section{Conclusion}
\label{Conclusion}
Summarizing, in this paper we have  studied the continuous thermodynamic phase transitions of AdS black holes according to the generalized Ehrenfest classification.  By using the Caputo fractional derivatives of thermodynamic potentials for both a charged black hole surrounded by quintessence,  $5D$ Gauss-Bonnet  and RN-AdS$_D$ black holes,  we find  that the fractional derivatives of the Gibbs free energy is always discontinuous at the critical point for $\beta = 4/3$ order, and diverges when $\beta>4/3$. These results suggest that  the $4/3$ order phase transition is robust and holds as far  one deals with static black hole with spherical symmetry. However, this feature is not universal and fails for axisymmetric solutions, as demonstrated for Kerr black hole where the phase transition happens at $\beta=1/3$ order. Nevertheless, further investigations of other black holes configurations are required to consolidate these findings, and establish a more involved classification of thermodynamic phase transitions.

\section*{Acknowledgments}
This manuscript has been released as a pre-print at arXiv database 
 \cite{Chabab:2020imt}.

\end{document}